\documentclass[12pt]{article}
\usepackage{latexsym,epsfig,graphics}
\newcommand{\be}{\begin{equation}}
\newcommand{\ee}{\end{equation}}
\newcommand{\bq}{\begin{eqnarray}}
\newcommand{\eq}{\end{eqnarray}}
\newcommand{\pigi}{\pi\,g^{2}}
\newcommand{\cues}{{\bf q}^{2}}
\newcommand{\cue}{{\bf q}}
\newcommand{\ips}{\int_{0}^{p^{+}}\! d\sigma\,}
\newcommand{\cuep}{{\bf q'}}
\newcommand{\bfv}{{\bf v}}
\begin{document}
\begin{titlepage}
\today          \hfill 
\begin{center}

\vskip .5in

{\large \bf More On The Connection Between Planar Field Theory And String Theory}
\footnote{Notice:* This manuscript has been authored by Korkut
Bardakci under Contract No. DE-AC02-05CH11231 with the U.S. Department of
Energy. The United States Goverment retains and the publisher, by
accepting the article for publication, acknowledges that the United
States Goverment retains a non-exclusive, paid-up, irrevocable, world-wide
license to publish or reproduce the published form of this manuscript,
or allow others to do so, for United States Goverment purposes.}
\vskip .50in


\vskip .5in
Korkut Bardakci
\footnote{Email: kbardakci@lbl.gov}
\vskip 9pt
{\em Department of Physics\\
University of California at Berkeley\\
   and\\
 Theoretical Physics Group\\
    Lawrence Berkeley National Laboratory\\
      University of California\\
    Berkeley, California 94720}
\end{center}

\vskip .5in

\begin{abstract}

\vskip 9pt

We continue work on the connection between world sheet representation
of the planar $\phi^{3}$ field theory and string formation. The present
article, like the earlier work, is based on the existence of a solitonic
solution on the world sheet, and on the  zero mode
fluctuations around this solution. The main advance made in this paper
is the removal of the cutoff and the transition to the continuum limit 
on the world sheet. The result is an action for the modes whose 
energies remain finite in this limit (light modes). The expansion of this
 action about a dense background of graphs on the world sheet leads to the
formation of a string. 

\end{abstract}
\end{titlepage}

\newpage
\renewcommand{\thepage}{\arabic{page}}
\setcounter{page}{1}
\noindent{\bf 1. Introduction}
 \vskip 9pt

This article is the continuation of the previous work on the same subject [1].
The basic problem, investigated in a series of papers [2,3,1], was to
develop a string formulation for the planar graphs of the
 $\phi^{3}$ field theory on the world
sheet. Both the present work and references [1,2] are based on the world sheet
picture developed in [3], which was in turn inspired by 't Hooft's seminal
work [4]. Our motivation for writing a paper which has quite a bit of
overlap with [1] is to clarify and extend the results obtained there, as well
as to make a correction. To make  the material intelligible to a
reader not familiar with the previous work,  we have inevitably to do
some reviewing. We will start by briefly reviewing the main
results of [1], and then summarize the advances made in this paper in
comparison to [1]. As we go along, we will also
 preview each section, and again stressing
the overlap and the differences with the previous work.

The starting point of reference [1] was the world sheet field theory
developed in [2]. In section 2, we review the world sheet description
of the planar graphs of the $\phi^{3}$ theory [3,4], and in section 3,
the world sheet field theory that reproduces these graphs [2] is
described. The  theory
is formulated in terms a complex scalar $\phi$ and a two component fermion
$\psi_{1,2}$; a central role is played by the world sheet field $\rho$,
which is
a composite of fermions (eq.(7)). Roughly speaking,
  $\rho$ measures the density
of graphs on the world sheet. An important question is whether $\rho_{0}$,
the average or the expectation value of  $\rho$, is different from zero.
 $\rho_{0}$ vanishes in any finite order of perturbation theory,
whereas a non-zero $\rho_{0}$ means that the world sheet is densely
covered by graphs, and the contribution of higher order graphs dominate.
In references [1] and [2], it was argued, on the basis of mean field
approximation, that the ground state of the model was realized in the
$\rho_{0}\neq 0$ phase.

In this paper, we will not investigate the
ground state of this unphysical model, which may not even exist.
 Instead, we will
have a more modest goal: Do the graphs that are dense on the world sheet
have a string description? It seems plausible that such set of graphs
would lead to a Nambu-Goto type action and in fact, this idea
motivated some of the early work on this subject [5,6].
To answer this question,  we will simply
 fix  $\rho_{0}$ at a non-vanishing value by introducing a
source coupled to $\rho$  and suitably tuning this source.
This is equivalent to choosing by hand a set of graphs whose
average density is  $\rho_{0}$ and investigating their properties.
Here we do not probe too deeply into the dynamics of the   $\phi^{3}$
theory; on the other hand, the results obtained are more robust, and
hopefully, they are more generally applicable, possibly  to more
physical models.

In section 4, which is a partial review of a corresponding section in
[1], the world sheet fermions are bosonized. In the fermionic picture,
$\rho$ takes on two values, 0 and 1, corresponding to the presence or
the absence of a line on the world sheet. In contrast,
 in the bosonic picture, $\rho$
is a continuous field with values ranging from 0 to 1. The bosonic
picture provides a very convenient setup for the mean field
approximation, developed in the following sections. This section also
includes a correction to [1]: In that reference, it was incorrectly
stated that $\rho$ is a non-dynamical time independent field. Here
we show that, on the contrary, $\rho$ is a time dependent dynamical field,
and  it plays an important role in the dynamics of the string.

A crucial feature of the world sheet field theory is that it has solitonic
classical solutions. In fact, the trivial vacuum of the theory is
unstable against tadpole emissions, and the model is stablized by 
introducing the solitonic background. These classical solutions were studied
in [1] and [3], and in section 5, we briefly review them. An unusual 
feature is that the corresponding classical energy is ultraviolet
divergent. The degree of divergence depends on number of dimensions; in
most of the paper, we will take the number of transverse dimensions D
to be two (4 space-time dimensions). This simplifies the treatment
considerably: The model is superrenormalizable, the classical energy
is log divergent, and the coupling constant is finite. This divergence
can be removed by mass renormalization. In [1], this renormalization
was carried out for a specific value of $\rho_{0}$ corresponding to a
candidate ground state; here, since $\rho_{0}$ is variable
at our disposal, we
have to  require renormalizibility for all $\rho_{0}$ in the physical range
$0\leq \rho_{0}\leq 1$. This is a much stronger condition, and it fixes
part of the world sheet action.

To see how this comes about, consider the interaction vertex where
 three propagators meet (Fig.3). The prefactor $1/(2 p^{+})$
in the propagator (eq.(1)) has to be attached to the vertex, and
there are various ways of doing this. In [1], these prefactors
were attached to the vertex symmetrically, as in eq.(11), but there
are also asymmetrical ways of doing it. So long as $\rho$ is a discrete
 variable with values 0 and 1, as in the original fermionic version,
all of these vertices yield the same result, but once $\rho$ becomes
a continuous variable, this is no longer true. In section 5, we show
that requiring renormalizibility for arbitrary $\rho$ fixes the vertex
completely (eq.(26)). In contrast to the symmetrical vertex used in [1],
 this corresponds to an asymmetrical vertex. Apart from this issue of
renormalization, its precise form will not matter for any of the
other results of this article.

As explained in sections 2 and 3, to have a well defined 
field theory, one of the coordinates of the world sheet, $\sigma$, is
discretized in steps of ``a''. This amounts to compactifying the light
cone coordinate $x^{-}$, and it was extensively used both in field
theory [7], and in string theory [8] and also M theory [9]. In section
6, we study in some detail the continuum limit $a\rightarrow 0$ on the world
sheet. This  limit  was not
investigated systematically in [1], and it is an important problem, since
 the world sheet field theory greatly simplifies and a string picture clearly
emerges only in this limit. An undesirable feature of the field
theory with discretized $\sigma$ was its non-locality in this 
coordinate. In the limit  $a\rightarrow 0$, the non-locality goes away, and
the result is a local field
theory on the world sheet, with an action with no higher than first
  order  $\sigma$ derivative of the fields. The derivation of a local
world sheet theory in the continuum limit is one of central results
of the present work that goes beyond reference [1].

 The continuum hamiltonian
can be written as a sum of two terms: One that is finite in the
$a\rightarrow 0$ limit and the other that blows up as $1/a^{2}$ (eq.(41)).
In section 7, we address the question of how to interpret this result.
At first sight, it may appear that the energies and therefore the masses
of various modes go to infinity. There are, however two exceptions to
this observation. With some natural scaling of parameters, the mass
of the $\rho$ stays finite, and there is also a transverse vector mode
$\bfv$, whose mass is vanishes. This is easy to understand: 
The model is invariant under the translation of the momentum $\cue$
(eqs.(2) and (15)), but this invariance is spontanously broken by the
presence of the soliton, and $\bfv$ is
 the corresponding Goldstone mode.
 This mode was discovered in [1] by expanding to quadratic order
about the classical background; here, using translation invariance,
we derive an exact result, valid to all orders.

We call the two modes discussed above the light modes, and we assume
all the other modes, whose masses go to infinity in the continuum
limit, decouple from the light modes. This hypothesis of the
decoupling of heavy modes is widely used both in field theory and
string theory. In section 8, using this hypothesis, we derive
  the continuum limit, resulting in
a world sheet action in terms of the light modes $\rho$ and $\bfv$
alone (eq.(57)), which is the main result of this work. All reference to
the cutoff ``a'' and the momentum $\cue$ has disappeared from the problem.
 Expanding
about a constant background $\rho= \rho_{0}$ results in a free massive 
$\rho$ and a free string action for $\bfv$.

In section 9, we extend these results to $D=4$ (5+1 dimensions). The main
complication is a cutoff dependent coupling constant, characteristic
of $\phi^{3}$ in 6 dimensions, which is asymptotically free. Despite this
complication, everything works out pretty much the
same as in $D=2$, and a string is formed from the light mode $\bfv$.
 The main difference is that the
dimensionless coupling
constant is replaced by a mass parameter, and the theory can be renormalized
by identifying this mass with the slope of the string. Finally, section 10
summarizes our conclusions. 

\vskip 9pt

\noindent{\bf 2. The World Sheet Picture}

\vskip 9pt
The planar graphs of $\phi^{3}$ in the light cone representation of
't Hooft [4] have a simple represantation. The world sheet is parametrized
by the coordinates
$$
\tau=x^{+}=(x^{0}+x^{1})/\sqrt{2},\,\,\,\sigma=p^{+}=(p^{0}+p^{1})/
\sqrt{2}.
$$
The coordinate $\tau$ will serve as the light cone time. 
A general plane graph is represented by a collection of horizontal solid
lines (Fig.1), where the n'th line carries a D dimensional transverse
momentum ${\bf q_{n}}$.
\begin{figure}[t]
\centerline{\epsfig{file=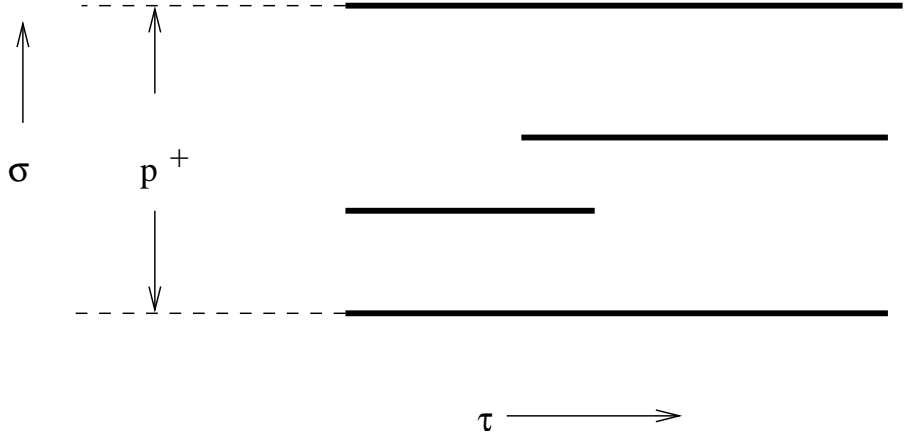, width=12cm}}
\caption{A Typical Graph}
\end{figure}
 Two adjacent solid lines labeled by n and n+1
correspond to the light cone propagator
\be
\Delta({\bf p_{n}})=\frac{\theta(\tau)}{2\,p^{+}}\,\exp\left(-i \tau\,
\frac{{\bf p_{n}}^{2}+ m^{2}}{2\,p^{+}}\right),
\ee
where ${\bf p_{n}}=\bf{q_{n}}- {\bf q_{n+1}}$ is the momentum flowing through
 the propagator. The interaction takes place at the beginning and at the end 
of each line , where a factor of g, the coupling constant is inserted.
Ultimately, one has to integrate over all possible locations and lengths
of solid lines, as well as over the momenta they carry.

We note that there is invariance under the translation
\be
{\bf q_{n}}\rightarrow {\bf q_{n}}+ {\bf r},
\ee
where ${\bf r}$ is a constant vector. This invariance will play an important
role in the subsequent development.

To avoid singular expressions later on, it is convenient to discretize
the coordinate $\sigma$ in steps of length a, which amounts to compactifying
the light cone coordinate $x^{-}=(x^{0} - x^{1})/\sqrt{2}$ at radius
$R= 1/a$. This type of compactification has proved useful in field
theory [7] and also in string theory [8], and in the M theory [9].
 In contrast, the time coordinate
$\tau$ will remain continuous. A useful way of visualizing the discretized
world sheet is pictured in Fig.2.
\begin{figure}[t]
\centerline{\epsfig{file=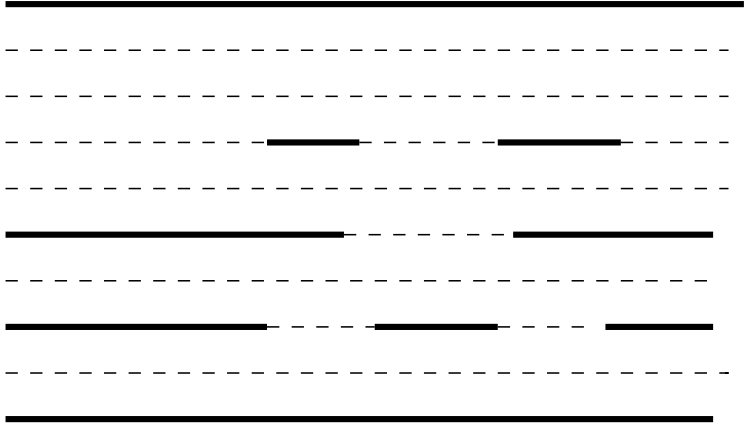, width=10cm}}
\caption{Solid And Dotted Lines}
\end{figure}
 The boundaries of the propagators are
marked by solid lines as before; but now, the bulk is filled by dotted
lines spaced a distance a apart.

We have also to specify the boundary conditions on the world  sheet. The
$\sigma$ coordinate is compactified by imposing periodic boundary
conditions at $\sigma=0$ and $\sigma= p^{+}$, where $p^{+}$ is the
total $+$ component of the momentum flowing the whole graph. This means
that the total transverse momentum ${\bf p}$ flowing through the graph is
zero. In
contrast, the boundary conditions at $\tau=\pm \infty$ will be left free.

\vskip 9pt

\noindent{\bf 3. The World Sheet Field Theory}

\vskip 9pt

In this section, we will briefly review the world sheet theory developed
in [1], which reproduces the light cone graphs described in the previous
section. We introduce the complex scalar field $\phi(\sigma,\tau,{\bf q})$
and its conjugate $\phi^{\dagger}$, which at time $\tau$ annihilate and
create a solid line carrying momentum ${\bf q}$ and located at site
labeled by $\sigma$.
 They satisfy the usual commutation
relations
\be
[\phi(\sigma,\tau,{\bf q}),\phi^{\dagger}(\sigma',\tau,{\bf q'})]
=\delta_{\sigma,\sigma'}\,\delta({\bf q}-{\bf q'}).
\ee
The vacuum, defined by
\be
\phi(\sigma,\tau,{\bf q})|0\rangle =0,
\ee
 corresponds to a state with only dotted lines (empty world sheet). By
applying $\phi^{\dagger}$'s on the vacuum, one can construct states with
arbitrary number of solid lines.

In addition, we will need a two component fermion field $\psi_{i}(\sigma,
\tau)$, $i=1,2$, and its adjoint $\bar{\psi}_{i}$, with the standard
anticommutation relations
\be
[\psi_{i}(\sigma,\tau),\bar{\psi}_{i'}(\sigma',\tau)]_{+} 
=\delta_{i,i'}\,\delta_{\sigma,\sigma'},
\ee
and propagate freely on an uniterrupted line. The fermion with $i=1$
lives on the dotted lines and the one with $i=2$ lives on the solid lines.
The main reason for introducing the fermions is to avoid unwanted
configurations on the world sheet. One type of such a configuration
corresponds to multiple solid lines at the same site, generated by repeated
applications of $\phi^{\dagger}$ at the same $\sigma$. To get rid of these
redundant configurations, we impose the constraint
\be
\int d{\bf q}\,\phi^{\dagger}(\sigma,{\bf q})\phi(\sigma,{\bf q})
=\rho(\sigma),
\ee
where
\be
\rho=\bar{\psi}_{2}\, \psi_{2},
\ee
which is equal to one on solid lines and zero on the dotted lines. This
constraint ensures that there is at most one solid line at each site.
Here and in the rest of the paper, it is understood that the fields
$\phi$, $\rho$ as well as the fermionic fields are functions of both
$\sigma$ and $\tau$, although for simplicity, we will frequently
not write down explicitly the dependence on one or both of these variables.

Fermions are also needed to avoid another set of unwanted configurations:
Propagators should be assigned only to adjacent solid lines; those
associated with non-adjacent solid lines are not allowed. For this
purpose, we define, for $\sigma_{j}>\sigma_{i}$
\be
\mathcal{E}(\sigma_{i},\sigma_{j})
=\prod_{k=i+1}^{k=j-1}(1-\rho(\sigma_{k})).
\ee
If $\sigma_{j}\leq \sigma_{i}$, $\mathcal{E}$ is defined to be zero.
The crucial property of this function  is that it
is equal to one only if the the two solid lines at $\sigma_{i}$ and
 $\sigma_{j}$ are seperated by dotted lines. If there are any solid
lines in between, it is zero. With the help of $\mathcal{E}$, the free
Hamiltonian (no interaction) can be written as
\bq
H_{0}&=&\frac{1}{2} \sum_{\sigma,\sigma'} \int d{\bf q} \int d{\bf q'}
\,\frac{\mathcal{E}(\sigma,\sigma')}{\sigma' -\sigma}\left(({\bf q} -
{\bf q'})^{2} +m^{2} \right)\nonumber\\
&\times& \phi^{\dagger}(\sigma, {\bf q})\phi(\sigma, {\bf q})\,
 \phi^{\dagger}(\sigma', {\bf q'})\phi(\sigma', {\bf q'})\nonumber\\
&+&\sum_{\sigma} \lambda(\sigma)\left(\int d{\bf q}\,  
\phi^{\dagger}(\sigma, {\bf q})\phi(\sigma, {\bf q}) -\rho(\sigma)
\right),
\eq
where $\lambda$ is a lagrangian multiplier enforcing the constraint (6).

It is now not too difficult to verify that, the operator
$\exp(-i\tau H_{0})$, applied on the states described following eq.(4),
reproduces a collection of free propagators given by (1), except for
the prefactor $1/(2 p^{+})$. The constraint (6)
eliminates unwanted states with multiple solid lines, and the presence
of $\mathcal{E}$ ensures that propagators are generated only by adjacent
solid lines.

Taking advantage of (6), it is possible to rewrite the free hamiltonian
in a simpler form:
\bq
H_{0}&=&\frac{1}{2}\sum_{\sigma,\sigma'} G(\sigma,\sigma')\Bigg(\frac{1}{2}
m^{2} \,\rho(\sigma) \rho(\sigma') +\rho(\sigma')\,\int d{\bf q}\,
{\bf q}^{2}\,\phi^{\dagger}(\sigma,{\bf q}) \phi(\sigma,{\bf q})
\nonumber\\
&-& \int d{\bf q} \int d{\bf q'} ({\bf q}\cdot {\bf q'})
 \phi^{\dagger}(\sigma, {\bf q})\phi(\sigma, {\bf q})\,
 \phi^{\dagger}(\sigma', {\bf q'})\phi(\sigma', {\bf q'})\Bigg)\nonumber\\
&+&\sum_{\sigma} \lambda(\sigma)\left(\int d{\bf q}\,  
\phi^{\dagger}(\sigma, {\bf q})\phi(\sigma, {\bf q}) -\rho(\sigma)
\right),
\eq
where we have defined
$$
G(\sigma,\sigma')=\frac{\mathcal{E}(\sigma,\sigma')+ \mathcal{E}
(\sigma',\sigma)}{|\sigma-\sigma'|}.
$$

It remains to introduce interaction. The two kinds of interaction
vertices are pictured in Fig.3.
\begin{figure}[t]
\centerline{\epsfig{file=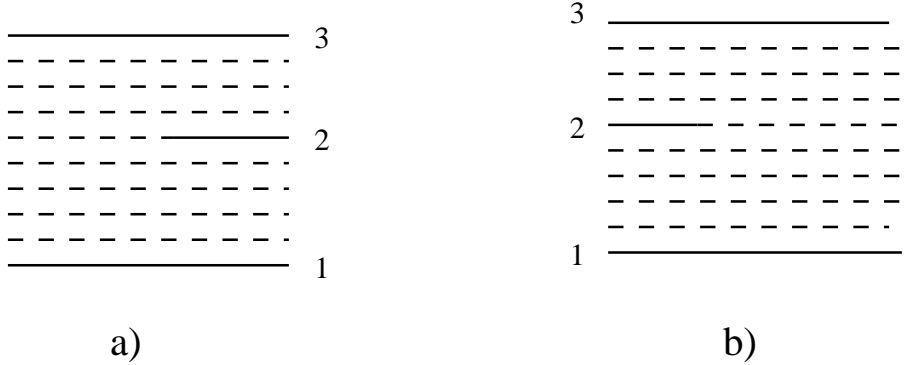, width=12cm}}
\caption{The Two $\phi^{3}$ Vertices}
\end{figure}
 In the first one, the operator
 $\phi^{\dagger}$ creates the solid line 2, and in the second figure,
$\phi$ destroys the solid line. This is, however, not the whole story; we
have to take care of  the missing factor of  $1/(2 p^{+})$ mentioned
above by attaching it to the vertices. There many ways of doing this; in
references [1,3], this was done by attaching a factor of
\be
V=\frac{1}{\sqrt{8\,p_{12}^{+}\,p_{23}^{+}\,p_{13}^{+}}} =
\frac{1}{\sqrt{8(\sigma_{2} -\sigma_{1})(\sigma_{3} -\sigma_{2})
(\sigma_{3} -\sigma_{1})}}
\ee
to each vertex. This is equivalent to splitting the factor $1/(2 p^{+})$
into two factors of $1/\sqrt{2 p^{+}}$ and attaching one to the
beginning and one to the end of the propagator. This is the most
symmetric assignment, but one could consider other ways of splitting this
factor, resulting in different expressions for $V$.
 Later on, we will choose to attach the whole factor to the
beginning (or to the end) of each propagator. Of course, in the exact theory,
all of these different assignments are equivalent; only when an approximation
is made, the differences between them emerge.  

Given a choice of $V$, we define
\be
W(\sigma_{2})=\sum_{\sigma_{1}<\sigma_{2}} \sum_{\sigma_{2}<\sigma_{3}}
\rho(\sigma_{1})\,\mathcal{E}(\sigma_{1}, \sigma_{3})\, \rho(\sigma_{3})
\,V(\sigma_{1}, \sigma_{2}, \sigma_{3}).
\ee
In this expression, the factor $\rho\,\mathcal{E}\,\rho$ multiplying $V$
picks the correct vertex configuration of Fig.3, and projects out the
unwanted configurations. With the help of $W$, the interaction hamiltonian
can be written as
\be
H_{I}=g\,\sum_{\sigma}\int d{\bf q}\,W(\sigma)\left(\phi(\sigma,{\bf q})\,
\rho_{+}(\sigma)+ \phi^{\dagger}(\sigma,{\bf q})\,\rho_{-}(\sigma)\right),
\ee
where $g$ is the coupling constant, and $\rho_{\pm}$ are defined by
$$
\rho_{+}= \bar{\psi}_{1} \psi_{2},\,\,\,\rho_{-}= \bar{\psi}_{2}
\psi_{1}.
$$
The additional factors of $\rho_{\pm}$ are needed to make sure that
a solid line is always paired with an $i=2$ fermion and a dotted line
with an $i=1$ fermion.

Finally, the total hamiltonian is given by
$$
H= H_{0} + H_{I},
$$
and the corresponding action by
\be
S=\int d\tau\left(\sum_{\sigma}\left(i\bar{\psi}\partial_{\tau}\psi
+ i \int d{\bf q}\,\phi^{\dagger}\partial_{\tau}\psi\right) -H(\tau)\right).
\ee

We note that the hamiltonian, as well as the action, is invariant under
the transformation
\be
\phi(\sigma,\tau, \cue)\rightarrow \phi(\sigma,\tau, \cue+ {\bf r}),
\ee
which is a restatement of (2) in the language of field theory. This invariance
will play an important role later on.

\vskip 9pt

\noindent{\bf 4. Bosonization}

\vskip 9pt

The hamiltonian defined by eqs.(10) and (13) is so far exact. It is also
non-local in the $\sigma$ direction and somewhat complicated. We will
 transform it into a local and simpler form later on.
As a first step, in preparation for the mean field approximation,
 it is very convenient to bosonize the fermions defined in the
previous section. This was done in  [1], and we follow this reference
closely. In addition to $\rho$ (eq.(7)), we introduce the bosonic field
$\xi$ and set,
\be
\rho_{+}=\bar{\psi}_{1} \psi_{2}= \sqrt{\rho -\rho^{2}}\,\,e^{i \xi}, \,\,\,
\rho_{-}= \bar{\psi}_{2} \psi_{1}=\sqrt{\rho -\rho^{2}}\,\,e^{- i \xi}.
\ee 
The kinetic energy term for $\psi$ in eq.(14) is then replaced by
\be
\int d\tau \sum_{\sigma} i\,\bar{\psi}\partial_{\tau}\psi \rightarrow
\int d\tau \sum_{\sigma} \xi\, \partial_{\tau} \rho.
\ee

One can check that that this action produces the correct equations of 
motion and the correct commutation relations for the fermionic 
bilinears.  It is therefore an exact transformation. However, we note
that bosonization has replaced  discrete variables by continuous
ones. For example, according to its original definition, $\rho$ could 
only take on the values 0 and 1, but as an independent bosonic field, it
can vary continuously between 0 and 1. It natural to interpret it as
the probability of finding a solid line at a given location. In going
over to the bosonic picture, the problem has been reformulated in terms
of a continuously valued field, but so long as the functional integral
over $\rho$ is carried out exactly, the reformulation is  exact.
The approximation comes later: Unable to do the functional integral
exactly, we will resort to 
  the mean field approximation, which amounts to evaluating the 
 integral  approximately
by expanding $\rho$ about a constant non-zero background $\rho_{0}$.
Since, by its definition,  $\rho_{0}$ measures the average density of
solid lines on the world sheet, if $\rho_{0}\neq 0$, the world sheet is
densely covered by solid lines, or equivalently, by Feynman graphs.
As explained in the introduction, we will arrange to have $\rho_{0}\neq 0$ 
by adding an external field $J$ coupled to $\rho$ to the action (eq.(44));
the minimum of the potential for $\rho$ will then be at an adjustable
$\rho=\rho_{0}$. 

It is possible to eliminate the field $\xi$ from the problem, leaving behind
only $\rho$. This is because the action is invariant under the
transformation
\bq
\phi&\rightarrow& \exp(-i \alpha)\,\phi,\,\,\, \phi^{\dagger}\rightarrow
\exp(i \alpha)\,\phi^{\dagger},\nonumber\\
\xi&\rightarrow& \xi +\alpha,\,\,\, \lambda\rightarrow \lambda -
\partial_{\tau}\alpha,
\eq
where $\alpha$ is an arbitrary function of $\sigma$ and $\tau$, so this
is a gauge transformation on the world sheet. We can therefore fix the gauge
by setting
$$
\xi=0.
$$
The equation of motion with respect to $\xi$ at $\xi=0$
should then be imposed as a constraint:
\be
\partial_{\tau} \rho -i g\,\sqrt{\rho-\rho^{2}}\,W\, \int d{\bf q}
\left(\phi -\phi^{\dagger}\right)=0.
\ee

 Using the equations
of motion, this  can be shown to be equivalent to
$$
\partial_{\tau}\left(\rho - \int d{\bf q}\, \phi^{\dagger}\,\phi
\right)=0.
$$
This nothing but the time derivative of the constraint (6), so everything
is consistent. In fact, this shows that (6) can be imposed as an initial
 condition, and it will then be preserved by the equations of motion.
Eq.(19) replaces eq.(22) of reference [1], which was incorrect.

After bosonization, the interaction hamiltonian becomes
\be
H_{I}\rightarrow g\,\sum_{\sigma} W(\sigma) \int d{\bf q}\,\left(
\phi(\sigma,{\bf q}) + \phi^{\dagger}(\sigma,{\bf q})\right),
\ee
where we have taken the liberty of absorbing the factor of
$\sqrt{\rho -\rho^{2}}$ into the definition of $W$.

\vskip 9pt

\noindent{\bf 5. The Classical Solution}

\vskip 9pt

In this section, we will treat $\rho(\sigma,\tau)$ and $\lambda(\sigma,\tau)$
as fixed classical external fields, with $\rho$ satisfying the condition
$$
0\leq \rho \leq 1.
$$
Of course, ultimately we have to integrate functionally over them,
but for the time being, they will be fixed.
 The function $W$, which can be expressed in terms of $\rho$,
also becomes an external field, and as a consequence, $H_{I}$, being linear
in $\phi$ and $\phi^{\dagger}$, becomes a tadpole term, and destabilizes
the vacuum of eq.(4).
 It is then
natural to eliminate this term by shifting $\phi$ by
\be
\phi= \phi_{0}+\phi_{1},\,\,\,\phi^{\dagger}=
\phi_{0}^{\star}+ \phi_{1}^{\dagger}.
\ee
$\phi_{0}$ is then a classical field,  determined by the tadpole
elimination condition, and $\phi_{1}$ represents quantum fluctuations
around this classical background. Although after this shift, $H_{I}$
disappears, interaction is now encoded in the expression for $\phi_{0}$,
which depends on the coupling constant $g$ (see eq.(30)).

We choose $\phi_{0}$ so that it depends on the vector ${\bf q}$ only
through ${\bf q}^{2}$ (rotation invariance).
This ensures that the last term on the right of eq.(10) does not contribute,
and we have  the simple problem of a hamiltonian with only
quadratic and linear terms. The tadpole elimination equation gives
\be
\phi_{0}=- g\,\frac{W}{E\,{\bf q}^{2} +\lambda},
\ee
where,
$$
E(\sigma)=\frac{1}{2} \sum_{\sigma'} G(\sigma,\sigma')\,\rho(\sigma').
$$

Although we have eliminated one set of tadpoles, a new term of the form
\be
\Delta S=
\int d\tau \int d{\bf q}\sum_{\sigma}\left(i\, (\phi_{1}^{\dagger}-
\phi_{1})\,\partial_{\tau}\phi_{0}\right)
\ee
is generated in the action. Being linear in $\phi_{1}$, this term is
also a tadpole. If we choose $\rho$ to be $\tau$ independent, $\phi_{0}$
will also be $\tau$ independent, and this term will vanish. In this
special case, $\phi_{0}$ is a classical solution to the equations of
motion (a soliton), which was  introduced in references [3] and [1].
For the time being, however, we will keep $\rho$ general and
 not suppress its $\tau$ dependence.
  We will later see that the resulting extra term is
actually needed to provide $\rho$ with a kinetic energy term.

The classical contribution to the action is of special interest. In
general, due to the integration over ${\bf q}$, it has an ultraviolet
divergence, which depends on the dimension of the space-time. In what
follows, we will specialize to $D=2$ (4 space-time dimensions), except
in section 9, when we will  consider $D=4$ (6 space-time dimensions).
At $D=2$, $\phi^{3}$ is super renormalizable, and only the mass term is
logarithmically divergent. In our treatment, this divergence shows up
in the classical contribution to the action:
\bq
S_{c}&=& S(\phi=\phi_{0})= g^{2}\,\int d\tau\sum_{\sigma}\left(\frac{\pi\,
W^{2}}{E} +\pi\,\int_{0}^{\Lambda^{2}} d\cues\,
 \frac{W^{2}}{E\,{\bf q}^{2} +\lambda}\right)
\nonumber\\
&=& g^{2}\,\pi\,\int d\tau \sum_{\sigma}\frac{W^{2}}{E}\left(1+
\ln\left(\frac{E\,\Lambda^{2}}{\lambda}\right)\right),
\eq
where $\Lambda$ is an ultraviolet cutoff. In a Lorentz invariant theory,
we expect this divergence to contribute only to the mass term in eq.(10),
 so it should be proportional to
\bq
S_{m}&=&- \int d\tau \sum_{\sigma,\sigma'}\left(\frac{1}{4} m^{2}\,
 G(\sigma,\sigma')\,\rho(\sigma)\,\rho(\sigma')\right)\nonumber\\
&=& - \frac{1}{2} m^{2}\,\int d\tau \sum_{\sigma} E(\sigma)\,\rho(\sigma).
\eq

Setting the cutoff term in $S_{c}$ proportional to $S_{m}$ gives
$$
\frac{W^{2}}{E}=const.\,E\,\rho.
$$
If this condition is satisfied, the divergence can be eliminated by a mass
 counter term. Setting the constant equal to one, which can always be
 arranged by a redefinition of the coupling constant g, we have the
defining equation
\be
W= E\,\sqrt{\rho}
\ee
for the hitherto unspecified vertex function W. From now on, we will
assume that the divergence has been eliminated by a counter term,
so that $S_{c}$ can be dropped if we replace m by $m_{r}$,
the renormalized mass.

We have derived this  condition by requiring renormalizability
for an arbitrary $\rho(\sigma,\tau)$. However, the same
result can also be derived by attaching the factors $1/(2 p^{+})$ (eq.(1))
always to the beginning (or always to the end) of each propagator. In
contrast, a different distribution of this factor, such as the one in
eq.(11), spoils renormalizability when the background $\rho$ is arbitrary.
 This is surprising, since in the exact
theory, it should not matter how this factor is split between the beginning 
and the end of the propagator. This is true if $\rho$ is built out of
fermions (eq.(7)) and therefore it is restricted to
take on only the values one and zero. 
 However, after bosonization, $\rho$ is a continuous variable, and different
splittings of the factor $1/(2 p^{+})$ give different results. We
then choose the splitting that leads to eq.(26) and thereby makes mass
renormalization possible. It should be stressed that
apart from renormalizability, subsequent results
 we are going to obtain, concerning, for example string formation,
are robust. They
do not depend on the detailed form of $W$,
 but only on its general features.

So far we have treated $\lambda$ as a fixed external field, same as
$\rho$. We are now ready to carry out the functional integration over
$\lambda$. We first set
\be
\lambda= \lambda_{0} +\lambda_{1},
\ee
where $\lambda_{0}$ is the classical part and $\lambda_{1}$ is the
quantum fluctuation to be integrated over. It is now natural to
redefine the classical solution $\phi_{0}$ (eq.(22)) by replacing $\lambda$
by its classical part $\lambda_{0}$:
\be
\phi_{0}\rightarrow - g\,\sqrt{\rho}\,\frac{E}{E\,{\bf q}^{2} +
\lambda_{0}}.
\ee
Also, eq.(26) for W was used.

 $\lambda_{0}$ is determined by the classical equation of motion
it satisfies, or equivalently, by directly imposing the constraint (6)
on the redefined $\phi_{0}$. The result is,
\be
\lambda_{0}=\pi\,g^{2}\,E,
\ee
and,
\be
\phi_{0}= -g\,\frac{\sqrt{\rho}}{\cues +\pigi}.
\ee

Making use of (27), (29) and (30),
the term proportional to $\lambda$ in (9) can now be rewritten as
\be
\sum_{\sigma}\lambda\left(\int d\cue\, \phi^{\dagger} \phi -\rho \right)
=\sum_{\sigma} \left(\lambda_{0}\,\int d\cue\,
\phi_{1}^{\dagger} \phi_{1} + \lambda_{1}\,\int d\cue \left(
\phi_{0}\,(\phi_{1} +\phi_{1}^{\dagger})+ \phi_{1}^{\dagger} \phi_{1}
\right)\right),
\ee
and the integration over $\lambda_{1}$ gives the constraint
\be
\int d\cue \left(
\phi_{0}\,(\phi_{1} +\phi_{1}^{\dagger})+ \phi_{1}^{\dagger} \phi_{1}
\right)=0.
\ee
Therefore, this term can be replaced by
$$
\sum_{\sigma}\lambda_{0}\,\int d\cue\, \phi_{1}^{\dagger} \phi_{1},
$$
plus the constraint (32).

Finally, putting everything together, the full hamiltonian is,
\bq
H&=&\frac{1}{2} \sum_{\sigma,\sigma'} G(\sigma,\sigma')
\Bigg(\frac{m_{r}^{2}}{2}\,
\rho(\sigma) \rho(\sigma')+\rho(\sigma')\,\int d\cue\,\cues
\,\left(\phi_{1}^{\dagger} \phi_{1}\right)_{\sigma,\cue}\nonumber\\
&-&\int d\cue \int d\cuep\,(\cue\cdot \cuep) \left(\phi_{0} (\phi_{1}
^{\dagger}+\phi_{1})+\phi_{1}^{\dagger} \phi_{1}\right)_{\sigma,\cue}
\, \left(\phi_{0} (\phi_{1}
^{\dagger}+\phi_{1})+\phi_{1}^{\dagger} \phi_{1}\right)_{\sigma',\cuep}
\Bigg)\nonumber\\
&+&\sum_{\sigma}\lambda_{0}(\sigma)\,\int d\cue\,\left(\phi_{1}^{\dagger}
\phi_{1}\right)_{\sigma,\cue},
\eq
supplemented by the constraint (32). As explained earlier, the classical
part of $H$ is absorbed into mass renormalization. Also since $\phi_{0}$
and $\lambda_{0}$ satisfy the equations of motion, linear terms in
$\phi_{1}$ and $\phi_{1}^{\dagger}$ are absent.

\vskip 9pt

\noindent{\bf 6. The Continuum Limit}

\vskip 9pt

In this section, we will take  the continuum limit on the world sheet,
by letting a, the spacing in the $\sigma$ direction go to zero. 
First consider  the kinetic energy term in the action (14):
\be
S_{k.e}=2\,\int d\tau\sum_{\sigma}\,\int d\cue\,\, \phi_{1,i}\,
\partial_{\tau}\phi_{1,r},
\ee
where $\phi_{r}$ and $\phi_{i}$ are the real and imaginary components
of $\phi$.
In making transition to the continuum limit by
\be
\sum_{\sigma}\rightarrow \frac{1}{a}\,\ips,
\ee
  an extra factor of $1/a$ is introduced, which
has to be scaled away in order to have a canonically normalized
kinetic energy term.  $\phi_{0}$, which is real and therefore
 part of $\phi_{r}$, is 
 independent of $a$ (eq.(30)). If we want the classical solution to
survive in $a\rightarrow 0$ limit,  $\phi_{r}$ cannot be scaled by
$a$. Therefore, the only remaining possibility is to scale $\phi_{i}$
by $a$. For convenience, we redefine $\phi_{i}$ by letting
$$
\phi_{i}\rightarrow a\,\phi_{i}
$$
and replace in (34) the sum over $\sigma$ by an integral:
\be
S_{k.e}= 2\,\int d\tau \ips \int d\cue\,\,\phi_{i}\,\partial_{
\tau}\phi_{r}=2\,\int d\tau \ips \int d\cue \, \phi_{1,i}\,
\partial_{\tau}(\phi_{0}+ \phi_{1,r}).
\ee
Notice that $\phi_{i}=\phi_{1,i}$ since $\phi_{0}$ is real.

Next we would like to derive the continuum limit of the hamiltonian (33).
It will consist of two terms: One proportional to $1/a^{2}$ the 
other independent of $a$,
\be
H= \frac{1}{a^{2}}\,H^{(1)}+ H^{(2)}.
\ee
Terms proportional to positive powers of $a$ will be dropped since
they vanish in the continuum limit. The main problem is the factor
$\mathcal{E}$ in the definition of $G$ 
(eq.(8)), which is so far defined only for a
 discretized $\sigma$ coordinate. Consider an interval of $\sigma$
much bigger than the lattice spacing a:
$$
\Delta\sigma=\sigma_{j} -\sigma_{i}\gg a,
$$
so that, as  $a\rightarrow 0$, $\Delta\sigma$ is kept fixed.
If $\rho$ is different from zero in this interval,
 $\mathcal{E}(\sigma_{i}, \sigma_{i}+\Delta\sigma)$, which has
$n$ factors of $1-\rho$ in its definition, where
$$
n=\frac{\Delta\sigma}{a}-1,
$$
rapidly goes to zero as  $a\rightarrow 0$, $n\rightarrow \infty$.
 Then, in this limit,
 assuming that $\rho\neq 0$, the sums over $\sigma$ and $\sigma'$ in
eq.(33) have contribution only from very small $|\sigma- \sigma'|$
of the order of $a$. We will call this phenomenon localization: As  lattice
spacing $a$ goes to zero, $\sigma\rightarrow \sigma'$, and the non-local
terms in the hamiltonian approach a local limit.  This result is
based on two assumptions:\\
a) Generically, $\rho$ is different from zero. Since, by its definition
(eq.(7)), $\rho$ measures the density of solid lines on the world sheet,
we are assuming that the world sheet is densely populated with Feynman
graphs. Later, we will justify this assumption by showing that one can
expand $\rho$ about a background $\rho_{0}\neq 0$.\\
b) $\phi$ and $\rho$ 
are slowly varying fields. More precisely, this means that
the typical length scale that governs the variation of these fields is not
$a$ but, as we shall see later, it is $g$, the coupling constant, which
 has the dimension of mass and is independent of $a$.

A typical term involving $G$ in  $H$ is of the form
\be
H\simeq \frac{1}{2}
\sum_{\sigma,\sigma'} G(\sigma,\sigma')\,A(\sigma)\,B(\sigma').
\ee
Under the assumption of localization, it is natural
to expand the factor $A\,B$ in powers of $\sigma'-\sigma$:
\be
A(\sigma)\,B(\sigma')=A(\sigma)\,B(\sigma)+ (\sigma'-\sigma) A(\sigma)\,
B''(\sigma) + \frac{1}{2} (\sigma' -\sigma)^{2}\,A(\sigma)\,B''(\sigma)
+ \cdots,
\ee
where primes indicate derivatives with respect to $\sigma$.
We have stopped at the second order in this expansion;
we shall shortly see that each power of $\sigma'-\sigma$ corresponds to
a power of $a$, and  terms higher than second order  
 do not contribute in the continuum limit. Plugging in the above 
expansion in (33), we encounter the sums
\bq
&&\sum_{\sigma'} G(\sigma,\sigma')=2 \sum_{n=0}^{\infty}\frac{(1-\rho)^{n}}
{a (n+1)}=\frac{2}{a}\, L(\sigma)\nonumber\\
&&\sum_{\sigma'} (\sigma' -\sigma)\,G(\sigma,\sigma')=0,\nonumber\\
&&\sum_{\sigma'} (\sigma' -\sigma)^{2}\,G(\sigma,\sigma')=
2 a\sum_{n=0}^{\infty} (n+1)\,(1-\rho)^{n}= \frac{2 a}{\rho^{2}},
\eq
where,
$$
L=-\frac{\ln(\rho(\sigma))}{1-\rho(\sigma)}.
$$

In carrying out these sums, we have taken $\rho$ to be $\sigma$
independent. Actually, a weaker condition of slowly varying
 $\rho(\sigma)$ is sufficient,
provided that we also expand $G$ up to second order in $\sigma' -\sigma$.
This amounts to expanding various fields up to second order in
derivatives with respect to $\sigma$ around a slowly varying background.
A typical term in
the hamiltonian  can then be cast into the following form:
\bq
H&\rightarrow&\ips\Bigg( \left(\frac{L(\sigma)}{a^{2}}+
f_{1}(\rho(\sigma))\,\left(\rho'(\sigma)\right)^{2}+
f_{2}(\rho(\sigma))\,\rho''(\sigma)\right)\,A(\sigma)
B(\sigma)\nonumber\\
&-& \frac{1}{2\,\rho^{2}(\sigma)}\,A'(\sigma)\,B'(\sigma)\Bigg),
\eq
where we have converted the sum over $\sigma$ into an integral by (35).
$f_{1,2}$ are local functions of $\rho$ which can be computed by
expanding $G$ in powers of $\sigma'-\sigma$. Since they will not be
needed in the subsequent development, we have not written them out
explicitly.

We note that $H$ has one term that goes as $1/a^{2}$, and other terms
that are $a$ independent. Also, it is local, with at most two
derivatives in $\sigma$ and in $\tau$. It is gratifying that
the $a\rightarrow 0$ limit produces a local action of the standard
canonical type on the world sheet.

 Eq.(40) can now be used directly in  the hamiltonian to
find the continuum limit. This results in a rather lengthy expression.
In the next section, we shall see that for the sector of the model we
are interested in, the light sector, there is a great simplification;
so instead, we will carry out the computation for only this sector.

\vskip 9pt

\noindent{\bf 7. The Light Sector Of The Model}

\vskip 9pt

A crucial feature of eq.(41)  is the first term, which
 blows up as $a\rightarrow 0$. Since the kinetic energy term (36) remains
finite in this limit, this implies that masses of various modes go to
infinity. As was done in reference [1], one can show this explicitly by
expanding to quadratic order around a fixed background in $\rho=\rho_{0}$.
 In any case, this result is pretty obvious
without any calculation. The important question is whether there are
any ``light'' modes, whose masses stay finite as $a\rightarrow 0$.
 In what
follows, we will indeed identify two such modes. We will then assume that
the ``heavy'' modes will decouple in continuum limit, and only the light
sector will remain behind. Although we have no rigorous proof of it,
it is a hypothesis very commonly used in field theory.

One of the two light modes is $\rho$ itself. Its mass
comes from the first term in eq.(33):
\be
H_{m\rho}=\frac{m_{r}^{2}}{4} \sum_{\sigma,\sigma'} G(\sigma,\sigma')\,
\rho(\sigma)\,\rho(\sigma')\rightarrow \frac{m_{r}^{2}}{2 a^{2}}
\ips L(\sigma)\,\rho^{2}(\sigma).
\ee
We have taken the continuum limit using (40) and kept the $1/a^{2}$ term.
The mass comes from expanding it around a fixed background
$\rho_{0}$. Clearly, to have a finite mass as $a\rightarrow 0$,
 $m_{r}$ must be scaled as
\be
m_{r}\rightarrow a\,u.
\ee
The parameter $u$ is dimensionless. This may seem surprising; however, 
 after the coefficient of the kinetic energy term is
normalized, the mass of  $\rho$ will turn out to be
  proportional to $g\, u$ (eq.(60)),
 which has the correct dimension since $g$ has dimension of mass.
The above scaling is therefore necessary to  have the mass term with the
correct dimension.
In contrast, $g$ needs no scaling: It already has the correct dimension
to be the coupling constant of the $\phi^{3}$ theory in four dimensions.

To finish this part of the story, we add an external source to $H_{m\rho}$,
which is really the potential term for $\rho$:
\be
H_{m\rho}\rightarrow \ips \left(-J\,\rho 
+\frac{1}{2} u^{2}\,L\,\rho^{2}\right),
\ee
where $L$ is given by (40), and $J$ is a $\sigma,\tau$ independent external
field (constant background). $J$ can be arranged to produce a minimum
at some  constant $\rho=\rho_{0}$, satisfying $0<\rho_{0}<1$,
  around which $\rho$ can be expanded by setting
\be
\rho=\rho_{0}+\rho_{1},
\ee
and expanding in $\rho_{1}$.

The second light mode is related to translation invariance in momentum
space (eq.(15)). This invariance is broken spontaneously by the classical
solution $\phi_{0}$, which is not translation invariant. This situation
is familiar from soliton and instanton physics; as a consequence of
Goldstone's theorem, a massless zero mode develops. Protected by Goldstone's
theorem, this mode stays massless also in the limit $a\rightarrow 0$,
and therefore it belongs to the light sector.
We will identify and quantize this mode by means of the collective
coordinate method, widely used for quantizing in a soliton background.

 Consider the field configuration
\be
\phi_{r}\rightarrow \phi_{0}(\cue+\bfv)=
- g\,\frac{\left(\rho(\sigma,\tau)\right)^{1/2}}
{\left(\cue+\bfv(\sigma,\tau)\right)^{2}+\pigi}.
\ee
What we have done is to promote the constant vector ${\bf r}$ into
the field $\bfv(\sigma,\tau)$ on the world sheet. In the action, we
will therefore make the ansatz
\be
\phi_{1,r}\rightarrow \phi_{0}(\cue+\bfv)-\phi_{0}(\cue).
\ee
On the other hand, we will keep $\phi_{1,i}$ arbitrary, since
 in the first order formalism for the action (eq.(36)),
$\phi_{r}$ and $\phi_{i}$ cannot be  simultaneously fixed.
 However, we note that since
$\phi_{i}$ has been scaled by a factor of $a$,  higher order
terms in $\phi_{i}$, unless accompanied by compansating factors of
$1/a$, can be dropped in the limit $a\rightarrow 0$.

We can now replace $\phi_{1,r}$ by (46), and then take the continuum
limit of H using (40). 
Instead, it is
possible to simplify the computation drastically
and also understand the simplicity
of the final result by the following
observation: Translation invariance (15) means that the resulting action
is invariant under the shift
\be
\bfv\rightarrow \bfv+ {\bf r},
\ee
where ${\bf r}$ is again a constant vector. This not only forbids a mass
term for $\bfv$, but also a term such as
$$
\left(\partial_{\sigma,\tau}\bfv\right)^{2}\,(\bfv)^{2}
$$
is not allowed.  To eliminate these terms,
 one can treat
$\bfv$, $\partial_{\sigma}\bfv$ and $\partial_{\tau}\bfv$ as independent
variables at  fixed $\sigma$ and $\tau$,
 and set $\bfv=0$, while keeping $\partial_{\sigma,\tau}\bfv$,
and also $\rho$ non-zero. Of course, a direct calculation would give
the same result, and what we are doing is to take a shortcut.

Let us now see what happens to some of these terms in (33) after this
simplification. Consider the third term on the right:
With the identification
\bq
{\bf A}=-{\bf B}&=&\int d\cue\,\cue\left(
\phi_{0}(\phi_{1}^{\dagger}+\phi_{1})+\phi_{1}^{\dagger}
 \phi_{1}\right)
=\int d{\cue}\,\cue \left( 
2 \phi_{0}\,\phi_{1,r}+ \phi_{1,r}^{2}+a^{2}\,\phi_{1,i}^{2}\right)
\nonumber\\
&\rightarrow&\int d{\cue}\,\cue
\left( \phi_{0}^{2}(\cue+\bfv) -\phi_{0}^{2}(\cue)+
a^{2}\,\phi_{1,i}^{2}\right),
\eq
we can apply eq.(40). The first term, 
with no $\sigma$ derivatives, vanishes:
$$
\phi_{0}^{2}(\cue+\bfv) -\phi_{0}^{2}(\cue)\rightarrow 0
$$
as $\bfv\rightarrow 0$. Also, $a^{2}\,\phi^{2}_{1,i}$ can be dropped as
 $a\rightarrow 0$.
 Therefore, only the last term in (40) with sigma
derivatives survives. In this term, as explained above, we 
set $\bfv=0$, while keeping $\partial_{\sigma}\bfv$ non-zero.
 With these simplifications, we have, in the limit $a\rightarrow 0$
and $\bfv\rightarrow 0$,
\be
\partial_{\sigma}{\bf A}\rightarrow 4 g^{2}\,\rho\,\int d\cue\,
\cue\,\frac{\cue \cdot \partial_{\sigma}\bfv}{\left(\cues+ \pigi
\right)^{3}}=\rho\,\partial_{\sigma}\bfv,
\ee
and therefore,
\be
\frac{1}{2}\sum_{\sigma,\sigma'} G(\sigma,\sigma') {\bf A}(\sigma)
\cdot {\bf B}(\sigma')\rightarrow \frac{1}{2} \ips \,
\left(\partial_{\sigma}\bfv\right)^{2}.
\ee

Now consider the second and the last terms on the right in (33). Both
of these terms contain the factor
\be
\phi_{1}^{\dagger} \phi_{1}\rightarrow \left(\phi_{0}(\cue+\bfv)
-\phi_{0}(\cue)\right)^{2}+ a^{2}\,\phi_{1,i}^{2}.
\ee
As $\bfv\rightarrow 0$, the first term vanishes. We cannot, however, drop the
second term as $a\rightarrow 0$, since there is a factor 
$1/a^{2}$ multiplying it. To see this, we rewrite these terms as
\bq
&&\frac{1}{2}\sum_{\sigma,\sigma'} G(\sigma,\sigma')\,\rho(\sigma')
\int d\cue\,\cues\,(\phi_{1}^{\dagger}\phi_{1})_{\sigma,\cue}
+\sum_{\sigma} \lambda_{0}(\sigma) \int d\cue\, (\phi_{1}^{\dagger}
\phi_{1})_{\sigma,\cue}\nonumber\\
&\rightarrow&\frac{1}{2} \sum_{\sigma,\sigma'} G(\sigma,\sigma')\,
\rho(\sigma')\,\int d\cue\,(\cues+\pigi)\,a^{2}\,\phi_{1,i}^{2}(
\sigma',\cue)\nonumber\\
&\rightarrow& \ips \int d\cue\, L(\sigma)\,\rho(\sigma)\,(\cues+\pigi)
\phi_{1,i}^{2}(\sigma,\cue),
\eq
where we have used,
$$
\lambda_{0}=\pi\,g^{2}\,E=
\frac{\pi\,g^{2}}{2}\sum_{\sigma'} G(\sigma,\sigma')\,\rho(\sigma')
\rightarrow \frac{\pi\,g^{2}}{a}\,L(\sigma)\,\rho(\sigma),
$$
which follows from (40). The above factor of $1/a$, combined with an
additional $1/a$ coming from converting the sum over $\sigma$ into
an integral, cancels the factor of $a^{2}$ multiplying $\phi_{1,i}^{2}$, so
 the final result is $a$ independent.

Putting together (51) and (53), the hamiltonian for the light sector is
\be 
H= H_{m\rho}+\ips \int d\cue\, L(\sigma)\,\rho(\sigma)\,(\cues+\pigi)
\phi_{1,i}^{2}(\sigma,\cue)+ \frac{1}{2} \ips \,
\left(\partial_{\sigma}\bfv\right)^{2}.
\ee
 Notice that the terms
proportional to $1/a^{2}$ in the hamiltonian have disappeared: the light
sector is really light.

The above hamiltonian has to be supplemented by the constraint (32).
 This presents no problem: in the
limit $a\rightarrow 0$, the ansatz (47) automatically satisfies this
constraint:
\be
\int d\cue \left(
\phi_{0}\,(\phi_{1} +\phi_{1}^{\dagger})+ \phi_{1}^{\dagger} \phi_{1}
\right)\stackrel{a\rightarrow 0}{\rightarrow} \int d\cue\left(
\phi_{0}^{2}(\cue+\bfv)- \phi_{0}^{2}(\cue)\right)=0.
\ee

One remaining question is whether there are additional light modes we
might have missed. In reference [1], by expanding
to quadratic order around a constant
background, it was shown that the masses of all the other modes
go as $1/a^{2}$ as $a\rightarrow 0$.

\vskip 9pt

\noindent{\bf 8. The Light Sector Action And String Formation}

\vskip 9pt

In the previous section, we presented the hamiltonian (54) for the light modes.
The corresponding action is given by (36). We need
\be
\partial_{\tau}\phi_{r}\rightarrow \partial_{\tau}\phi_{0}( 
\sigma,\tau,\cue+\bfv)\stackrel{\bfv\rightarrow 0}{\rightarrow}
 -\frac{g\,\partial_{\tau}\rho}
{2 \sqrt{\rho}\left(\cues+\pigi\right)}+ \frac{2 g\, \sqrt{\rho}\,
\cue\cdot \partial_{\tau}\bfv}{\left(\cues+\pigi\right)^{2}}.
\ee
The resulting expression is  at most quadratic in $\phi_{1,i}$.
This field can then be eliminated by doing the gaussian functional integral
over it, which amounts to solving its equation of motion.
 The result is
\bq
S&=&\int d\tau \ips\left(
 \int d\cue\,\frac{g^{2}\left(\frac{2 \rho^{1/2}\,\cue
\cdot \partial_{\tau}\bfv}{(\cues+\pigi)^{2}}-\frac{\partial_{\tau}
\rho}{2 \rho^{1/2}\,(\cues+\pigi)}\right)^{2}}{\rho\,L\,(\cues+\pigi)}
-\frac{1}{2}
u^{2}\,\rho^{2}\,L+J\,\rho\right) \nonumber\\
&=&\int d\tau \ips\left(\frac{(\partial_{\tau}\bfv)^{2}}{6 \pi^{2}\,
g^{4}\,L} -\frac{1}{2} (\partial_{\sigma}\bfv)^{2} +
\frac{(\partial_{\tau}\rho)^{2}}{8\pi\,g^{2}\rho^{2}\,L}-\frac{1}{2}
u^{2}\,\rho^{2}\,L+J\,\rho\right).\nonumber\\
&&
\eq

The above world sheet action is the fundamental result of the
present paper. There is no longer any reference to the target space
except for the number of components of $\bfv$;
 the field $\phi$, as well as the momentum $\cue$
and the cutoff ``a'' have all disappeared from the problem. This great
simplification is the result of the decoupling of heavy modes and of
localization on the world sheet in the limit $a\rightarrow 0$.

In deriving this result, so far there have been no approximations.
The only hypothesis that was needed was the decoupling of the heavy
states in the limit $a\rightarrow 0$. However, this action, which
has inverse powers of $\rho$, is
meaningful only if it is expanded around a no-zero value of $\rho$. The
natural choice for this constant $\rho_{0}$ is the minimum of the potential
in eq.(44). This expansion also coincides with mean field
approximation. The
quadratic terms in this expansion are of particular interest:
\bq
S_{2}&=&\int d\tau \ips\left(\frac{(\partial_{\tau}\bfv)^{2}}{6 \pi^{2}\,
g^{4}\,L(\rho_{0})} -\frac{1}{2} (\partial_{\sigma}\bfv)^{2} +
\frac{(\partial_{\tau}\rho_{1})^{2}}{8\pi\,
g^{2}\rho_{0}^{2}\,L(\rho_{0})}-\frac{1}{2}\,z^{2}(\rho_{0})\,
\rho_{1}^{2}\right)\nonumber\\
&\rightarrow&\int d\tau \ips \Bigg(\frac{1}{2}(\partial_{\tau}
\tilde{\bfv})^{2}-3\,\pi^{2}\,g^{4}\,L(\rho_{0})\,(\partial_{\sigma}
\tilde{\bfv})^{2}+ \frac{1}{2} (\partial_{\tau}\tilde{\rho}_{1})^{2}
\nonumber\\
&-&\frac{1}{2}\,\mu^{2}(\rho_{0})\,\tilde{\rho}_{1}^{2}\Bigg),
\eq
where we have scaled the fields by
\be
\bfv= \sqrt{3\,\pi^{2}\,g^{4}\,L}\,\,\tilde{\bfv},\,\,\,
\rho_{1}=\sqrt{4\,\pi\,g^{2}\,\rho_{0}^{2}\,L}\,\,\tilde{\rho}_{1}.
\ee

This is the world sheet action for a
 massive field $\tilde{\rho_{1}}$ plus a string generated by $\tilde{\bfv}$.
The mass $\mu$ of $\tilde{\rho_{1}}$ is given by
\be
\mu^{2}=4\,\pi\,g^{2}\,u^{2}\,L(\rho_{0})\,
\rho_{0}^{2}\,z^{2},\,\,\,
z^{2}=\frac{3- 4\rho_{0}+\rho_{0}^{2}+ 2\,
\ln(\rho_{0})}{2 (\rho_{0}-1)^{3}}.
\ee
$\mu^{2}$ is positive in the physical range $0<\rho_{0}<1$. The string slope
is 
\be
\alpha'=\left(12\,\pi^{4}\,g^{4}\,L(\rho_{0})\right)^{1/2}.
\ee
The dimensions of both the $\rho_{1}$ mass and the string slope are
provided by the coupling constant $g$, the only dimensional parameter in
the problem in the absence of $a$.

We note that, if we set $\rho_{1}=0$, we end up with a standard free
string in the light cone frame [10]. This is because
 translation invariance (48) does not
allow any higher order interaction terms  involving $\bfv$ alone.
 These non-linear terms
are however induced through the coupling between $\rho$ and $\bfv$
in the full action. Also, although there is no $\partial_{\sigma}\rho$
term in the action, it will be generated as a higher order contribution.
To avoid confusion, we emphasize that since we are summing  the 
planar graphs, the string is always a free string. String interactions can 
only come from  the non-planar graphs. What is happening is that with
$\rho$ couplings included, the string becomes non-linear: The straight
trajectories of the linear string are replaced by curved trajectories.

Superficially, it might appear that the expansion discussed above is
the same as
expanding in powers of $g$. Howevever, this is not quite right;
in  the limit
$g\rightarrow 0$, both the  string slope and  the
$\rho$ mass vanish. From the string perspective, this is a singular
limit, so it is better to trade $g$ for the string slope and give up
on expanding in powers of $g$. What we have here are two complementary
pictures: In the limit $g\rightarrow 0$, the appropriate picture is
the usual field theory picture of graph expansion. String picture
makes sense only for a world sheet densely covered by graphs, which
is clearly requies  $g\neq 0$, since otherwise  higher order
graphs needed to cover the world sheet would be suppressed.

\vskip 9pt

\noindent{\bf 9. Four Transverse Dimensions}

\vskip 9pt

In this section, we will   discuss the $D=4$ case, mostly
pointing out the changes that have to be made in going from $D=2$
to $D=4$. The
development all the way to section 5, eq.(23), is independent of the
number of dimensions and remains unchanged. The first change occurs
in the calculation of the classical action $S_{c}$ (eq.(24)), which is
now quadratically divergent:
\bq
S_{c}&\simeq& - g \int d\tau\sum_{\sigma}W(\sigma,\tau)\int d{\cue}\,
\phi_{0}(\sigma,\tau,\cue)=g^{2}\int d{\tau} \sum_{\sigma}\int d\cue\,
\frac{W^{2}}{E\,\cues+\lambda}\nonumber\\
&\simeq& g^{2}\,\pi^{2}\,\Lambda^{2}
\int d\tau \sum_{\sigma} \frac{W^{2}}{E}.
\eq
Here $\Lambda$ is the ultraviolet cutoff and we have kept  the 
leading quadratic divergence. The only difference between this and
 eq.(24) is that a log divergence is replaced by a quadratic one.
 As before, requiring it be proportional to the mass term in (33) for
renormalizability,  we again get
exactly eq.(26) for W. Also splitting $\lambda$ as in eq.(27),  eq.(28) for
$\phi_{0}$ remains unchanged.

To determine $\lambda_{0}$, we again impose the constraint (6),
which now involves a logarithmically divergent integral. The result is
\be
\frac{1}{g_{0}^{2}}=\pi^{2}\,\int_{0}^{\Lambda^{2}} d\cues \,\frac{
\cues}{(\cues+\lambda_{0}/E)^{2}}
= \pi^{2}\,\left(\ln\left(\frac{E}{\lambda_{0}}\,
\Lambda^{2}\right)-1\right),
\ee
where we have replaced $g$ by $g_{0}$ to make it clear that
we are dealing with the bare coupling constant. Since  $g_{0}$
is $\sigma,\tau$ independent, it follows that
\be
\lambda_{0}(\sigma,\tau)=\mu_{0}^{2}\,E(\sigma,\tau),
\ee
where $\mu_{0}$ is a $\sigma,\tau$ independent constant and eq.(63) can
be rewritten as
\be
\frac{1}{g_{0}^{2}}= \pi^{2}\,\left(
\ln\left(\Lambda^{2}/\mu_{0}^{2}\right)-1\right).
\ee
We will determine $\mu_{0}$ later on in terms of the slope of the string;
all we need to know at the moment is that
$$
\Lambda/\mu_{0}\rightarrow \infty
$$
as $\Lambda\rightarrow \infty$. Putting together these changes,
we have the result
\be
\phi_{0}= - g_{0}\,\frac{\sqrt{\rho}}{\cues+\mu_{0}^{2}}.
\ee

From this point on, till eq.(50) in section 7, there are no significant
changes except for the replacement
$$
\pigi\rightarrow \mu_{0}^{2}
$$
in the expression for $\phi_{0}$. However, in eq.(50), the integral
is over four dimensions and it is logarithmically divergent. This
equation now reads
\bq
\partial_{\sigma}{\bf A}&\rightarrow& 4 g^{2}\,\rho\,\int d\cue\,
\cue\,\frac{\cue \cdot \partial_{\sigma}\bfv}{\left(\cues+ \mu_{0}^{2}
\right)^{3}}\nonumber\\
&=& g_{0}^{2}\,\pi^{2}\,\ln\left(\frac{\Lambda^{2}}{\mu_{0}^{2}}\right)
\,\partial_{\sigma}\bfv\rightarrow \rho\,\partial_{\sigma}\bfv.
\eq
In the last step, using eq.(65) for $g_{0}$, the log divergence
cancelled. The final result is identical to $D=2$ case, and eq.(51)
remains unchanged.

Next, we turn to the eq.(57). The coefficient of $(\partial_{\sigma}\bfv)^{2}$
is unchanged, but the coefficients of $(\partial_{\tau}\bfv)^{2}$ and
$(\partial_{\tau}\rho)^{2}$ will be different. The differences come from
 the replacement $\pigi\rightarrow \mu_{0}^{2}$ and from the change 
 of the dimension of integration.  The integrals in
(57) are convergent at $D=4$, and the resulting action is
\bq
S&=&\int d\tau \ips \Bigg(\frac{\pi^{2}\,g_{0}^{2}}{12 L\,\mu_{0}^{4}}\,
(\partial_{\tau}\bfv)^{2} -\frac{1}{2} (\partial_{\sigma}\bfv)^{2}+
\frac{\pi^{2}\,g_{0}^{2}}{8 \rho^{2}\,L\,\mu_{0}^{2}}\,
(\partial_{\tau}\rho)^{2}\nonumber\\
&-&\frac{1}{2} u^{2}\,\rho^{2}\,L+J\,\rho\Bigg).
\eq
Expanding around $\rho=\rho_{0}$ as before, we again get a linear string plus
a massive $\rho$ field. The string slope is given by
$$
\alpha'=\left(\frac{24 \,L(\rho_{0})\,\mu_{0}^{4}}{g_{0}^{2}}
\right)^{1/2},
$$  
or,
\be
\frac{\mu_{0}^{2}}{\alpha'}=\left(24 \pi^{2}\,L(\rho_{0})\left(
\ln\left(\Lambda^{2}/\alpha'\right)-\ln\left(\mu_{0}^{2}/\alpha'\right)
\right)\right)^{-1/2}.
\ee

The model is renormalized by keeping the physical parameter $\alpha'$
fixed as  $\Lambda\rightarrow\infty$, and solving for $\mu_{0}$ and
$g_{0}$ in terms of $\alpha'$ and $\lambda$.
 To leading order in $\Lambda$,
$$
\frac{\mu_{0}^{2}}{\alpha'}
\rightarrow \left(24\,\pi^{2}\,L(\rho_{0})
\,\ln\left(\Lambda^{2}/\alpha'\right)
\right)^{- 1/2},
$$
and,
\be
\frac{1}{g_{0}^{2}}\rightarrow \pi^{2}\,\ln\left(\Lambda^{2}/\alpha'\right),
\ee
as $\Lambda\rightarrow \infty$. $\phi^{3}$ is known to be asymptotically
free in $5+1$ dimensions, and
 this dependence of the bare coupling constant on the cutoff is
characteristic of such a theory. We note that the contribution of
$\mu_{0}^{2}$ to $1/g_{0}^{2}$ goes as $\ln(\ln\Lambda))$ and it is 
negligible  compared to the leading $\ln(\Lambda)$ dependence.

So, despite the complications caused by the cutoff dependent coupling
constant, the string sector at $D=4$ works out pretty much as at $D=2$
after slope renormalization. The $\rho$ sector of the model is, however,
more problematic. In the limit of large $\Lambda$, the coefficient
of the kinetic energy term for $\rho$ goes to zero as
$$
\frac{g_{0}^{2}}{\mu_{0}^{2}}\rightarrow \left(\ln\left(\Lambda^{2}/
\alpha'\right)\right)^{-1/2}.
$$
Again expanding about $\rho=\rho_{0}$ and keeping only quadratic terms, we 
have a simple harmonic oscillator at each value of $\sigma$, decoupled
from the others. For the usual SHO lagrangian
$$
\mathcal{L}=\frac{1}{2}\,m\,\dot{x}^{2}- \frac{1}{2}\,k\,x^{2},
$$
the limit we are considering corresponds to $m\rightarrow 0$. In the
zero mass limit, the energies of all the excited states go to
infinity, and after subtracting the zero point energy, we are left
with only the ground state at zero energy. Therefore, $\rho_{1}$
collapses into time independent projection operators:
$$
\rho_{1}(\sigma)\rightarrow |0\rangle_{\sigma}\,\langle 0|_{\sigma},
$$
where $|0\rangle_{\sigma}$ are ground states, one for each $\sigma$,
that are mutually orthogonal. Further investigation, which we will
not undertake here, is necessary to verify that the interaction terms
do not change this picture.

The foregoing discussion is incomplete in one important respect:
we have tacitly assumed that the one loop result (70) for the coupling
constant holds all the way from energies of the order of $\Lambda$
to the string slope $\alpha'$. Of course, what is needed is a
renormalization group analysis for the running coupling constant.
We leave this interesting problem for future research.

\vskip 9pt

\noindent{\bf 10. Conclusions}

\vskip 9pt

The main result of the present article is the derivation of world sheet
action given by eqs. (57) and (68). The main advance made on reference [1]
is that these actions are now defined on a continuously parametrized
world sheet and the cutoff parameter ``a'' which discretized the world sheet
has disappeared. There is also now an additional dynamical variable $\rho$;
expanding $\rho$ around the minimum of its potential, we recover a
linear string, but there are also fluctuations around this minimum which
signal deviations from a linear string.

There are still important problems left for future research. One of
them is the question of Lorentz invariance \footnote{See [11] for
an investigation of renormalization and Lorentz invariance in the
light cone formulation.}. Although we have not done so here, it is not
difficult to establish invariance under the so called light cone Galilean
subgroup of the Lorentz group, but full Lorentz invariance is still
an open question. Another important target for research is to extend
world sheet methods to more physical theories. Although some initial
attempts were made in this direction [12,13],  much still remains
to be done.

\vskip 9pt

\noindent{\bf Acknowledgement}

\vskip 9pt

This work was supported in part by the Director, Office of Science,
Office of High Energy Physics of the U.S. Department of Energy under the
 Contract DE-AC02-05CH11231.

\newpage

{\bf References}

\vskip 9pt

\begin{enumerate}

\item K.Bardakci, JHEP {\bf 0903} (2009) 088, arXiv:0901.0949.
\item K.Bardakci, JHEP {\bf 0810} (2008) 056, arXiv:0808.2959.
\item K.Bardakci and C.B.Thorn, Nucl.Phys. {\bf B 626}(2002) 287,
hep-th/0110301.
\item G.'t Hooft, Nucl.Phys. {\bf B 72} (1974) 461.
\item H.P.Nielsen and P.Olesen, Phys.Lett. {\bf B 32} (1970) 203.
\item B.Sakita and M.A.Virasoro, Phs.Rev.Lett. {\bf 24} (1970) 1146.
\item A.Casher, Phys.Rev. {\bf D 14} (1976) 452.
\item R.Giles and C.B.Thorn, Phys.Rev. {\bf D 16} (1977) 366.
\item T.banks, W.Fischler, S.H.Shenker and L.Susskind, Phys.Rev.
{\bf D 55} (1997) 5112, hep-th/9610043.
\item P.Goddard, J.Goldstone, C.Rebbi and C.B.Thorn, Nucl.Phys.
{\bf B 56} (1973) 109.
\item C.B.Thorn, Nucl.Phys. {\bf B 699} 427, hep-th/0405018,
D.Chakrabarti, J.Qiu and C.B.Thorn, Phs.Rev. {\bf D 74} (2006) 045018,
hep-th/0602026.
\item C.B.Thorn, Nucl.Phys. {\bf B 637} (2002) 272, hep-th/0203167,
S.Gudmundsson, C.B.Thorn and T.A.Tran, Nucl.Phys. {\bf B 649} 3-38,
hep-th/0209102.
\item C.B.Thorn and T.A.Tran, Nucl.Phys. {\bf B 677} (2004) 289,
hep-th/0307203.

\end{enumerate}

\end{document}